\documentclass[twocolumn,twoside,letterpaper,11pt]{article}
\usepackage{graphicx,xrrc,natbib}
\usepackage{times}

\begin{document}


\title{JET/ENVIRONMENT INTERACTIONS OF FR-I AND FR-II RADIO GALAXIES}



%
%
%
%


\author{    J.H. Croston, M.J. Hardcastle, M. Birkinshaw, D.M. Worrall}  
\institute{ Astrophysics Group, University of Bristol         } 
\address{   H.H. Wills Physics Laboratory, Tyndall Avenue, Bristol, BS8 1EJ, UK        } 
\email{     judith.croston@bris.ac.uk, m.hardcastle@bris.ac.uk,
  mark.birkinshaw@bris.ac.uk, d.worrall@bris.ac.uk} 


\maketitle

\abstract{We present the results of new XMM observations of FR-I and
FR-II radio galaxies that show the importance of jet/environment
interactions for both radio-source structure and the properties of the
surrounding gas. The FR-I observations reveal that the distribution of
hot gas determines the radio-lobe morphology, and provide evidence that
the subsonic expansion of the lobes heats the surrounding gas. The
FR-II observations show that the sources are at equipartition and in
pressure balance with their surroundings. Finally, we present an
analysis of a sample of elliptical-dominated groups, showing that
radio-source heating is common.}

\section{Introduction}

Interactions between radio galaxies and their surrounding
X-ray-emitting hot gas regulate the expansion of the radio plasma and
transfer large amounts of energy to the gas environment. X-ray
observations of radio-galaxy environments are therefore important both
for developing our understanding of radio-galaxy properties and
dynamics, and for studying the impact of radio galaxies on groups and
clusters.

In the early Universe, radio galaxies and quasars are a possible
source of the energy input needed to reconcile cold dark matter models
of structure formation and the observed properties of clusters (e.g.
Sanderson et al. 2003, Roychowdhury et al. 2004). It has been argued
for some time that the energy injected by AGN into cluster gas is
likely to be important in balancing radiative cooling in cluster cores
(e.g. Binney \& Tabor, 1995), and it is now increasingly accepted that
the energy input of radio galaxies can reconcile X-ray observations
with cooling-flow models (Churazov et al. 2002, Fabian et al. 2003).
X-ray observations can provide information about the heating mechanism
used to transfer energy from the radio source to the gas; for example,
{\it XMM-Newton} and {\it Chandra} observations of the nearest radio
galaxy Centaurus A (Kraft et al. 2003, and these proceedings; Worrall
et al. these proceedings) have shown the importance of shock-heating
by young sources. Insight into other heating mechanisms has also come
from X-ray observations, such as those of ``ripples'' in the Perseus
cluster (Fabian et al. 2003).

Both low-power (FR-I) and powerful (FR-II) radio galaxies are expected
to heat their environments. FR-I radio galaxies are thought to expand
subsonically for most of their lives, so that they should be in
approximate pressure balance with their surroundings; they will do
$P$d$V$ work on the gas as they expand, but we would not expect
shock-heating, except in young sources. In contrast, the standard
model of FR-II expansion is that the jets inflate a supersonically
expanding cocoon of material that shocks and compresses the
surrounding gas. In this model, we might expect to be able to observe
a rim of hotter gas around the radio source. X-ray observations can
test the validity of these models, as they constrain properties of the
radio source, such as particle content, magnetic field strength and
internal pressure, and of the environment, e.g. density and pressure.

In the following sections we describe our {\it XMM-Newton}
observations of FR-I and FR-II radio galaxies. Details of these
observations and further discussion are presented in Croston et al.
(2003, 2004). We also discuss a survey of {\it ROSAT}-observed groups
that provides evidence that radio-source heating is common.

\section{Observations of FR-I radio galaxies}

\begin{figure*}
\centering{\vbox{
\includegraphics[width=12.0cm]{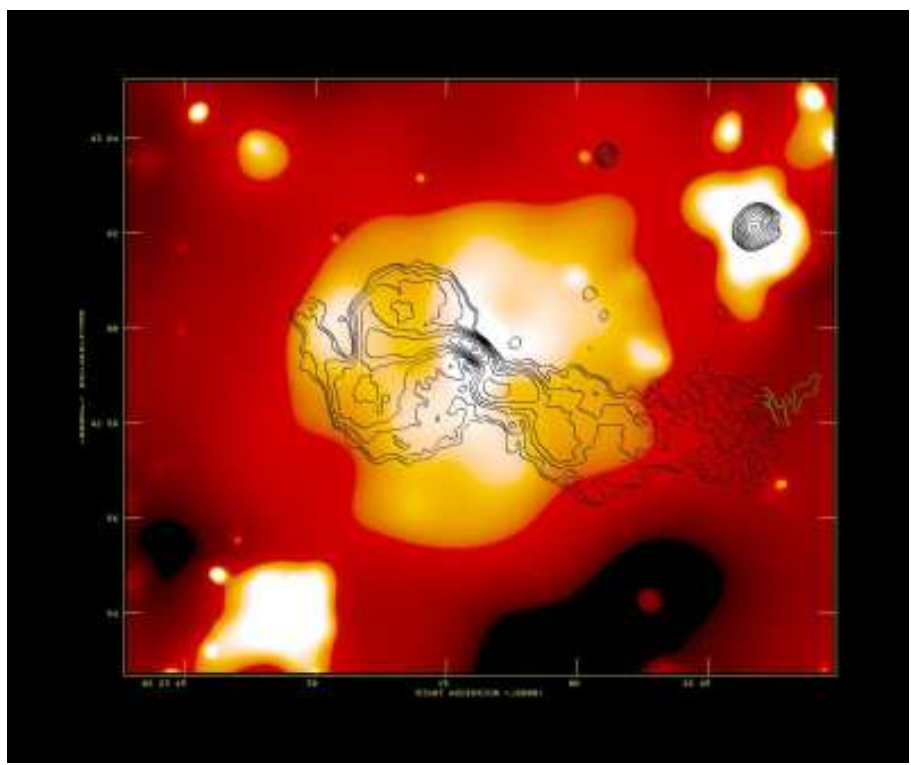}
\vskip 10pt
\includegraphics[width=8.5cm]{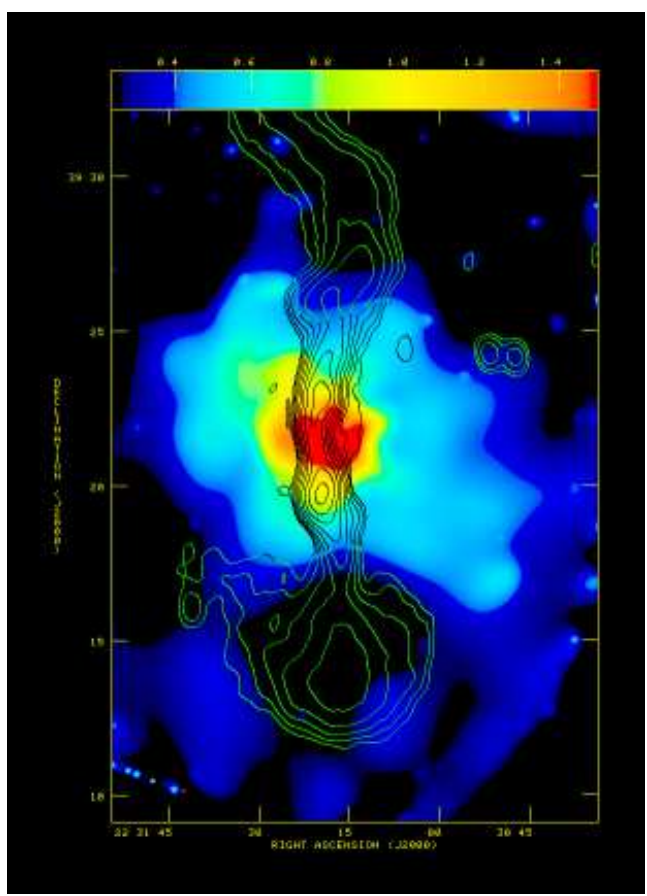}}}
\caption{The X-ray environments of the FR-I radio galaxies 3C~66B
  (top) and 3C~449 (bottom). Colour is the smoothed, background
  point-source subtracted, vignetting corrected, combined MOS1, MOS2
  and pn {\it XMM-Newton} data. Contours for 3C~66B are from the 1.4-GHz radio
  map of Hardcastle et al. (1996), and for 3C~449 are from a 608-MHz
  radio map from the 3CRR atlas (Leahy, Bridle \& Strom 1998).}
\label{fr1s}
\end{figure*}

We observed the two radio galaxies 3C~66B ($z=0.0215$) and 3C~449
($z=0.0171$) with {\it XMM-Newton} for 20 ks each. Fig.~\ref{fr1s}
shows the X-ray emission measured by the three {\it XMM} cameras
with radio contours overlaid, for both radio galaxies. Below
we discuss what the observations reveal about the role of environment
in shaping the radio lobes, the radio-source particle content and
dynamics, and the impact of the radio galaxies on their environment.

\subsection{Environmental influence on lobe morphology}

The two sources were chosen due to their asymmetrical radio
morphologies, so that we could investigate whether the asymmetries
could be explained by environmental conditions rather than intrinsic
differences in the jets. Fig.~\ref{fr1s} shows clear evidence that
jet/environment interactions are taking place. There are deficits in
X-ray surface brightness at the positions of the radio lobes, which
show that the lobes are displacing material. More importantly, there
is a clear relationship between the distribution of X-ray-emitting gas
and the shapes of the radio lobes. For both sources, the round,
sharply bounded lobe (3C~66B east and 3C~449 south) is embedded in
bright X-ray emitting material, whereas the narrower plumes (3C~66B
west and 3C~449 north) are relatively free of surrounding material.
This suggests that the rounded lobes have formed where the jets flowed
into dense material, and expansion along the jet axis was slowed;
conversely, such lobe structures failed to form on the opposite side
of the sources, because the jets encountered less dense material and
continued to flow more freely.

More evidence that the environment has played an important role in
determining the source structure comes from the bright blob of
emission at the end of the eastern jet of 3C~66B. This denser gas
appears to be blocking the jet expansion, leading to the sharp
boundary at the end of the lobe. It also appears that material in the
south is beginning to flow around the obstacle. In Section 2.3, we
discuss evidence that this blob of gas is being heated by its
interaction with the jet.

\subsection{Radio-source dynamics and particle content}

It has been known for some time that the minimum internal pressures of
FR-I radio lobes are typically lower by at least an order of magnitude
than external pressures inferred from X-ray observations (e.g.
Morganti et al. 1998; Worrall \& Birkinshaw 2000). We measured the
external pressures in the environments of 3C~66B and 3C~449 and
compared them with the minimum internal lobe pressures obtained from
radio data. Both sources appear to be underpressured by factors of
$\sim 20$, consistent with the earlier work on this type of source.
This discrepancy can be solved either by relaxing the minimum energy
assumption or by assuming some additional non-radiating particle
component in the lobes that is providing the required pressure [see
Hardcastle, Worrall \& Birkinshaw (1998) for more discussion of this
problem].

Our X-ray observations allow us to rule out the possibility of
increasing the relativistic electron population (and decreasing the
magnetic field strength) to increase the internal pressure, because an
electron population sufficiently large to provide pressure balance
would produce X-ray inverse-Compton emission at an easily detectable
level, which we do not see. Therefore, if a departure from
equipartition solves the pressure balance problem, then it must be in
the direction of magnetic dominance.

We can also rule out a dominant pressure contribution from entrained
gas at the temperature of the surroundings, as the observed level of
thermal emission from the lobe regions is too low. We cannot rule out
a contribution from relativistic protons, but the required
proton-to-electron number ratios are high ($\sim200$). An additional
population of low-energy electrons is also possible; however, the
required electron spectrum is physically implausible. We conclude that
a contribution from significantly heated, entrained gas is perhaps the
most likely explanation.

\subsection{Radio-source heating}

The bright blob of gas that is interacting with 3C~66B's eastern jet
is hotter than the surrounding material ($kT_{{\rm blob}} = 2.4\pm0.4$
keV; $kT_{{\rm env}} = 1.73\pm0.03$ keV). We interpret this
temperature difference as evidence for local heating due to the work
being done by the jet on the gas. In addition to this local heating,
there is also evidence for more general heating by 3C~66B. The average
group temperature of $1.73$ keV is significantly hotter than predicted
by L$_{X}$/T$_{X}$ relations for groups. This result is discussed
further in Section 4.

\section{Observations of FR-IIs}

\begin{figure*}
\centering{\vbox{
\includegraphics[width=10cm]{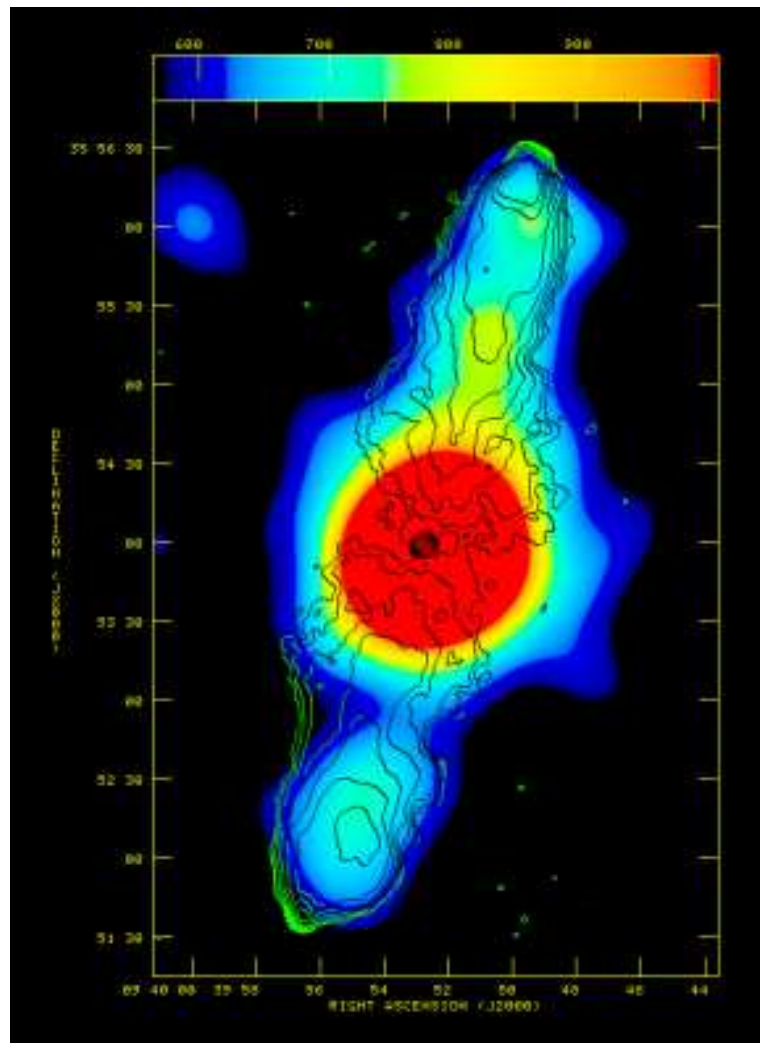}
\vskip 10pt
\includegraphics[width=14.0cm]{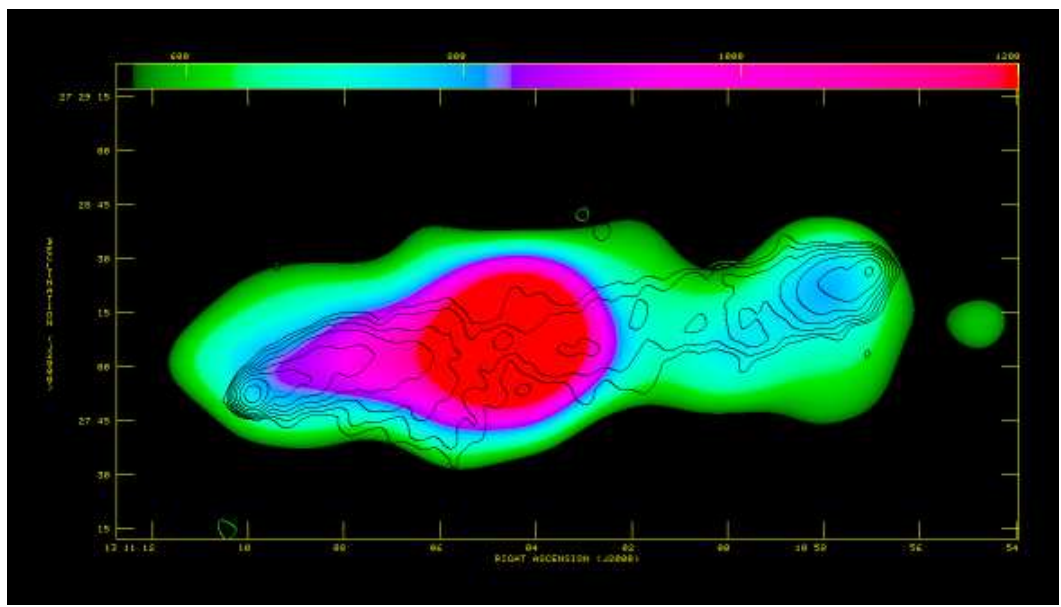}}}
\caption{The X-ray environments of the FR-II radio galaxies 3C~223
  (top) and 3C~284 (bottom). Colour is the smoothed, background
  point-source subtracted, combined MOS1, MOS2 and pn {\it XMM-Newton}
  data. Contours for 3C~223 are from the 1.4-GHz radio map of
  Leahy \& Perley (1991), and for 3C~284 are from a radio map made from 1.4-GHz
  archive VLA data.}
\label{fr2s}
\end{figure*}

We observed the powerful radio galaxies 3C~223 ($z=0.1368$) and 3C~284
($z=0.2394$) with {\it XMM-Newton} for 35 ks and 45 ks,
respectively. Fig.~\ref{fr2s} shows the X-ray emission from the three
{\it XMM} cameras, with radio contours overlaid, for both
sources. Below we discuss the origins of the different components of
X-ray emission, and the implications for radio-source properties and
dynamics.

\subsection{Emission from the cores, lobes and environments}

Fig.~\ref{fr2s} shows that X-ray emission is detected from the cores,
lobes and hot-gas environments of both sources. The nuclear emission
is not discussed here -- for its interpretation, see Croston et al.
(2004).

The X-ray spectra for the lobe-related regions of both sources are
consistent with inverse-Compton emission, as in other FR-II radio
galaxies (e.g. Hardcastle et al. 2002; Isobe et al. 2002; Belsole et
al. 2004 and these proceedings). We used multi-frequency radio data to
model the electron population in the lobes of each source (using the
code of Hardcastle, Birkinshaw \& Worrall 1998), assuming that the
lobes are in the minimum energy condition, and determined the
predicted X-ray flux from inverse-Compton emission. For the lobes of
both sources (each modelled separately), the measured X-ray fluxes are
in good agreement with the predictions for IC scattering of the CMB,
with magnetic field strengths within a factor of 2 of equipartition.

Both sources have group-scale environments (L$_{X}$ $\sim 10^{43}$
ergs s$^{-1}$) with temperatures of $\sim1.5$ keV (3C~223) and $\sim1$ keV
(3C~284). These results are in agreement with the predictions of
L$_{X}$/T$_{X}$ relations. The data quality is not good enough to
measure small temperature variations such as local heating around the
radio source. In the following Section, we discuss the
implications of our results for the radio-source dynamics and heating.

\subsection{Pressure balance and dynamics of FR-IIs}

From our model for the lobe-related X-ray emission of 3C~223 and 3C~284
we find the internal energy of the radio lobes to be close to the
minimum energy values. We can compare these values with the external
pressure from the X-ray-emitting gas to study the lobe dynamics. We
find that the lobes of both sources appear to be in pressure balance
with their surroundings.

This result leads to a self-consistent picture for the properties of
these two large FR-II radio galaxies: they are at equipartition and in
approximate pressure balance. This is consistent with earlier work
(Hardcastle et al. 2002). If this picture is correct, it seems likely
that many moderately large FR-IIs of similar radio structure have
reached a stage where lateral expansion is no longer supersonic, so
that the cocoon will have collapsed in the centre, where the gas
pressure is highest (this is consistent with the ``pinched'' central
radio structure of 3C~223, 3C~284 and other sources). In this stage of
evolution, heating effects will move outwards at the sound speed and
potentially affect gas at greater distances from the radio lobes than
is possible in the supersonic stage where the heated material is
confined to a rim surrounding the source.

If, instead, we believe that these sources must be expanding
supersonically, then additional material is required in the lobes to
make them overpressured. Entrainment is not thought to be important in
FR-IIs, so that such additional material would have to have a
different origin from that in FR-Is (see 2.2). The supersonic cocoon
model for FR-IIs therefore requires that the agreement between the
detected X-ray emission and the equipartition prediction in both
sources is a coincidence: the energetics of both FR-Is and FR-IIs
would have to be dominated by particles other than the relativistic
electrons, with a different origin for the dominant particle
population in each type of source. We therefore conclude that the
self-consistent model of FR-II radio lobes at equipartition and in
pressure balance with their surroundings is more plausible.

\section{Evidence for radio-source heating in a sample of groups}

\begin{figure*}
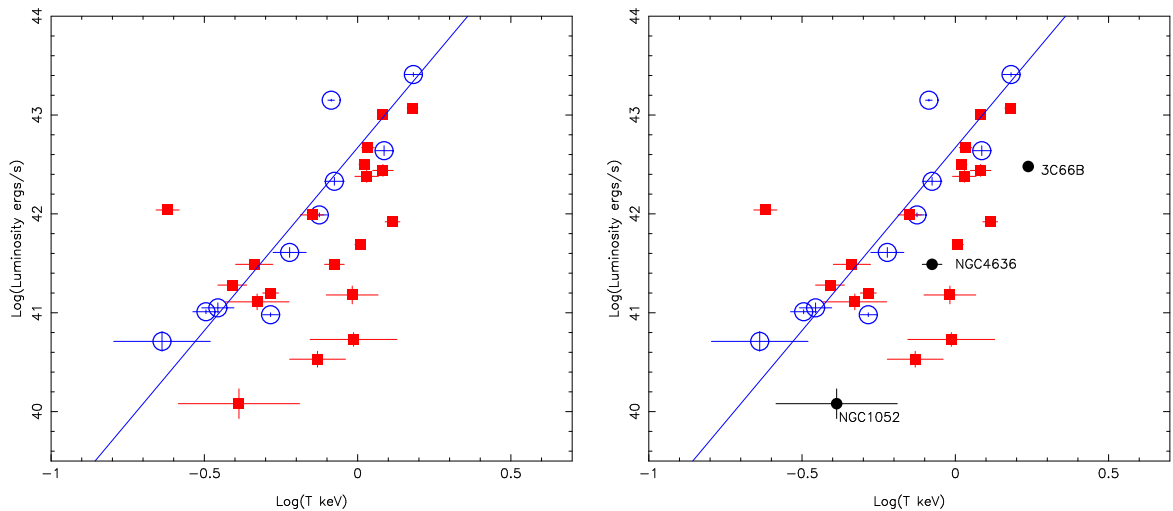

\centering{\hbox{
\includegraphics[width=7.5cm]{lt.ps}
\hskip 10pt
\includegraphics[width=7.5cm]{lt2.ps}}}
\caption{L$_{X}$/T$_{X}$ plots for ``radio-loud'' and ``radio-quiet''
  groups. On the left, the two subsamples are plotted, showing that
  the ``radio-loud'' groups (red filled squares) are hotter for their
  luminosity than ``radio-quiet'' groups (blue hollow circles). The
  best-fitting relation for ``radio-quiet'' groups is shown for
  comparison. The right-hand plot shows the same sample with the
  three groups that have additional evidence for heating marked by
  black filled circles.}
\label{lt}
\end{figure*}

Radio-loud AGN are preferentially found in groups or poor clusters
(Best 2004). In addition to studies of individual objects, it is
important to get an idea of how common radio-source heating is in
groups, the degree to which groups are affected by the presence of a
radio galaxy, and the length of time for which these effects persist.
We therefore looked at a sample of groups observed with {\it ROSAT} to
compare the gas properties of ``radio-loud'' and ``radio-quiet''
groups. Preliminary work on a small sample was described by Croston et
al. (2003). We discuss below more detailed analysis of a larger sample.

\subsection{Gas properties of groups with and without radio sources}

The sample used for this study is taken from Osmond \& Ponman (2004).
We used their measurements of gas temperature, luminosity and other
properties. We used DSS images to identify spiral-dominated groups and
excluded these from the analysis so as to obtain a sample of groups
homogeneous in their likelihood of containing a radio galaxy. We then
searched for associated radio sources using NED, NVSS and FIRST, and
confirmed identifications with group member galaxies using the DSS
images. Radio sources were found in 18 of the 29 groups in the sample.

We then divided the sample into radio-loud and radio-quiet groups
based on a radio luminosity cutoff, the choice of which does not
significantly affect our results. Fig.~\ref{lt} shows the two
subsamples plotted in L$_{X}$/T$_{X}$ space. It is immediately
apparent that the properties of the two samples differ. The
best-fitting L$_{X}$/T$_{X}$ relation for the ``radio-quiet''
subsample is shown on the plot. ``Radio-loud'' groups are more likely
to be to the right of the line, or hotter for a given luminosity. On a
K-S test, we find there is a less than 5 percent probability that the
two subsamples have the same parent population. There are two possible
interpretations of this result: radio-loud groups either have a
temperature excess or a luminosity deficit. In the next section, we
discuss the arguments in favour of each of these scenarios.

\subsection{Are the ``radio-loud'' groups heated?}

It is tempting to interpret this result as evidence that radio-source
heating is occurring in many groups. However, radio galaxies could
also affect the group luminosity. Radio galaxies must displace large
amounts of gas (as shown in Section 1), and this could have a
significant effect on the luminosity. For 3C~66B, we calculate that
the gas with which the radio source can have directly interacted
provides only 7 percent of the group's luminosity. It is therefore
unlikely that any removal of gas by the radio galaxy could produce the
luminosity deficits needed by this model. It is possible, however,
that the energy input from the radio galaxy could cause the group to
expand, which might have a larger effect. However, in some cases the
luminosity must be reduced by an order of magnitude, which seems
implausible.

There is another reason to favour a heating interpretation: several
individual sources showing a ``temperature excess'' do in fact have
additional evidence for radio-source heating. {\it Chandra}
observations of two groups in the ``radio-loud'' sample that are
hotter than predicted for their luminosity reveal detailed structure
in the group gas. The observation of the NGC~4636 group shows arms of
bright gas, which are hotter than the surrounding group (Jones et al.
2002), supporting a model where the departure from the L$_{X}$/T$_{X}$
relation is due to heating. This group is particularly interesting,
because the currently active radio source is too small and weak to
provide the required heating, showing that heating effects can be
longer lived than the radio galaxy. A short observation of the young
radio-source NGC~1052 reveals radio-related X-ray structure (Kadler et
al. 2003) suggestive of shock-heating similar to that seen in Cen A.
Finally, as reported in Section 1, the environment of 3C~66B is
significantly hotter than predicted, and there is other evidence for
radio-source heating in the group gas. These three sources are plotted
in the right-hand panel of Fig.~\ref{lt}. As there is direct evidence
for radio-source heating in three groups with a ``temperature
excess'', we conclude that this effect is largely due to heating
rather than a decrease in luminosity caused by the radio source.

\subsection{A more complicated picture?}

The three groups with direct evidence for radio-source heating
discussed in the previous section each represent different stages in
the life of a group containing a radio source. NGC~1052 is an example
of a young radio source likely to be shock-heating its surroundings
(like the inner regions of Cen A); 3C~66B is an older radio source
that is gently transferring energy to the gas; and NGC~4636 is a group
where the heating effects of a previous generation of radio source are
still important. These examples suggest that we would not expect all
groups containing a radio source to have similar gas properties.

In Fig.~\ref{lt}, we can see that there are some ``radio-loud'' groups
that do not show a temperature excess, and one particularly anomalous
group that is much cooler than predicted (NGC 3557). Although 3C~66B
would be capable of providing the energy needed to heat its
environment by the amount observed, the majority of the currently
active radio sources in the sample would not. We therefore conclude
that the effects of radio-source heating are long-lived (as shown by
the example of NGC~4636), so that much of the heating in the sources
far from the radio-quiet L$_{X}$/T$_{X}$ relation has been caused by
previous generations of radio-galaxy activity.

\section{Conclusions}

Our X-ray observations of FR-I radio galaxies have shown that:
\begin{itemize}
\item differences in the distribution of hot gas around the source lead to the
  development of the varied types of radio-lobe morphologies seen in
  FR-Is.
\item FR-Is cannot be in the minimum energy condition, and the
  additional pressure required for pressure balance with the
  surrounding gas cannot come from an increase in the density of
  synchrotron-emitting electrons or from entrained gas at the
  temperature of the surroundings. Heated, entrained material is
  perhaps the most plausible candidate to provide the required
  pressure.
\item FR-Is can heat their environments gently via subsonic expansion.
\end{itemize}

Our X-ray observations of FR-II radio galaxies have shown that:
\begin{itemize}
\item 3C~223 and 3C~284 are in the minimum energy condition and in
  pressure balance with their surroundings.
\item Older FR-IIs are likely to come into pressure balance and no longer shock-heat
  their surroundings via a supersonically expanding cocoon.  
\end{itemize}

We have shown that by studying the group atmospheres of individual
sources, we can investigate the energy transfer mechanisms for
different types of source and at different stages in source evolution.
The presence of shock-heating in sources like Cen A (Kraft et al.
2003) and more gentle heating in 3C~66B are consistent with a change
of energy transfer mechanism in FR-Is as they expand and come into
pressure balance. Our FR-II observations suggest that a similar
picture is also true for these objects, although the overpressured,
supersonic stage probably lasts for a larger fraction of the source's
life.

Finally, our study of a sample of groups has shown that radio-source
heating is common, and that the effects of radio-source heating on
groups are significant and long-lived.

\section*{Acknowledgments}

JHC acknowledges support from a PPARC studentship. MJH thanks the Royal
Society for a fellowship. We thank Trevor Ponman for providing
results in advance of publication.

\end{document}